\documentclass[aps,prd,notitlepage,reprint,onecolumn,showpacs,superscriptaddress,longbibliography,nofootinbib,floatfix]{revtex4-1}

\usepackage{amsmath,amssymb,amsfonts}
\usepackage{graphicx}
\usepackage{verbatim}
\usepackage{pifont}

\usepackage{hyperref}
\usepackage{slashed}
\usepackage{verbatim}
\newcommand{\be}{\begin{equation}}
\newcommand{\ee}{\end{equation}}
\newcommand{\bi}{\begin{itemize}}
\newcommand{\ei}{\end{itemize}}
\newcommand{\bea}{\begin{eqnarray}}
\newcommand{\eea}{\end{eqnarray}}

\newcommand{\ud}{\mathrm{d}}

\newcommand{\LCm}{{\scriptscriptstyle -}} 
\newcommand{\LCp}{{\scriptscriptstyle +}}

\newcommand{\LCperp}{{\scriptscriptstyle \perp}}

\usepackage[T1]{fontenc} 

\begin{document}

\title{The worldline approach to helicity flip in plane waves}

\author{Anton Ilderton}
\email[]{anton.ilderton@chalmers.se}
\affiliation{Department of Physics, Chalmers University, SE-41296 Gothenburg, Sweden}

\author{Greger Torgrimsson}
\email[]{greger.torgrimsson@chalmers.se}
\affiliation{Department of Physics, Chalmers University, SE-41296 Gothenburg, Sweden}

\begin{abstract}
We apply worldline methods to the study of vacuum polarisation effects in plane wave backgrounds, in both scalar and spinor QED. We calculate helicity-flip probabilities to one loop order and treated exactly in the background field, and provide a toolkit of methods for use in investigations of higher-order processes. We also discuss the connections between the worldline, S-matrix, and lightfront approaches to vacuum polarisation effects.
\end{abstract}
%
%
\maketitle
\section{Introduction}
Vacuum polarisation effects may be probed using strong electromagnetic fields, as may be generated by using e.g.~intense lasers. As such the measurement of ``vacuum birefringence''~\cite{Toll:1952}  (a matterless analogue of birefringence in optics) in the collision of x-ray and optical laser pulses, has been selected as a Flagship experiment at DESY~\cite{HIBEF}, following~\cite{Heinzl:2006xc}. For both the theoretical and experimental status of birefringence studies see the recent reviews~\cite{King:2015tba,HP}.

Vacuum polarisation effects stem of course from photon-photon scattering~\cite{Halpern:1934} and are captured, to lowest order, by the one-loop polarisation tensor in a background field. It is usually enough, considering current experimental abilities, to consider photon-photon interactions at the level of the low-energy Euler-Heisenberg effective action~\cite{Euler:1935zz,Heisenberg:1935qt}. As access to new parameter regimes becomes available~\cite{ELI}, it will though become increasingly important to go beyond low-energy approximations. Investigations in this direction are driven not only by phenomenological interest, but also by a desire to better understand higher-order, all-orders, and nonperturbative strong field effects in quantum field theory~\cite{Huet:2010nt,Shore:2007um,Dunne:2008kc,Huet:2011kd,Bakker:2013cea}. On this note, it is possible to calculate the polarisation tensor exactly for certain background fields. One example is a plane wave, as shown some years ago for a constant `crossed' field in~\cite{Narozhny:1968}, and for arbitrary plane wave shape and strength in~\cite{Becker-Mitter} (using Green's function methods) and~\cite{Baier:1975ff} (using operator methods). The polarisation tensor in both plane waves and magnetic fields has recently been reconsidered by several groups~\cite{Meuren:2013oya,Karbstein:2013ufa,Dinu:2013gaa,Meuren:2014uia,Karbstein:2015cpa}. This has lead to many new insights, in particular with regard to realistic field geometries~\cite{Gies:2013yxa,Dinu:2014tsa,Gies:2014wsa}, which must be accounted for in light-by-light scattering experiments~\cite{Lundstrom:2005za,King:2012aw,Gies:2013yxa} and vacuum birefringence experiments~\cite{Heinzl:2006xc,HIBEF,HP,Karbstein:2015xra}.

The original derivations of polarisation tensor results are rather involved, and it would be preferable to have more transparent expressions in order to improve our understanding of strong-field vacuum polarisation effects: even investigations of the plane wave case can reveal insights which may be extended to more general field configurations~\cite{Dinu:2014tsa,Torgrimsson:2014sra}. For these reasons we will here reconsider the plane wave polarisation tensor in another formalism, namely that of worldline path integrals~\cite{Feynman:1950ir,Feynman:1951gn,Halpern:1976gd,Halpern:1977he,Halpern:1977ru,Bern:1991aq,Strassler:1992zr} -- see \cite{CS} for an introduction and \cite{Schubert:2001he} for a review. The worldline formalism has proven powerful for the study of many topics, only a few examples of which are pair production~\cite{Affleck:1981bma,Dunne:2005sx,Dunne:2006st,Dunne:2006ur,CPL,Dumlu:2011cc,Strobel:2013vza,Ilderton:2014mla,Schneider:2014mla,Linder:2015vta,Ilderton:2015lsa,Ilderton:2015qda,Dumlu:2015paa},  photon splitting \cite{Adler:1996cja}, QCD~\cite{Ahmadiniaz:2015xoa}, string theories with contact interactions~\cite{Edwards:2014cga,Edwards:2014xfa}, and two loop Euler-Heisenberg effective action \cite{Fliegner:1997ra,Kors:1998ew}. Worldline path integrals also lend themselves to numerical evaluation using Monte Carlo methods~\cite{Gies:2001zp,Gies:2005bz,Gies:2011he,Schafer:2015wta}.

Our modest aim here is to use the worldline formalism to recover vacuum polarisation effects in inhomogeneous plane wave backgrounds, of arbitrary strength and shape. (This formalism has previously been applied to the study of the polarisation tensor in constant fields~\cite{Schubert:2000yt} and in inhomogeneous magnetic fields~\cite{Gies:2011he}).  As part of this calculation we will produce a new toolbox which can be used to apply worldline methods to more complicated higher-order strong-field processes in plane wave backgrounds.

This paper is organised as follows. Directly below we introduce our notation and conventions, and then describe the observables of interest, namely the helicity-flip probabilities in our chosen background field. In Section~\ref{SECT:SQED} we perform the worldline calculation of the helicity flip probability. In Section~\ref{SECT:QED} we extend this to the spinor case and compare two different worldline methods for calculating the spin contribution.  We discuss the results and conclude in Section~\ref{SECT:CONCS}. The appendix contains normalisations for worldline path integrals.
\subsection{Notation and conventions}
%
%
Our background field depends only on the phase $\phi:=n.x$ where $n^2=0$, and is transverse. We can always choose coordinates such that $n.x= x^0 + x^3 = x^\LCp$, `lightfront time', and then $\bf E$ and $\bf B$ point in the $\perp := \{x^1,x^2\}$ directions. We define $x^\LCm = x^0-x^3$. We use lightfront gauge $n.A=0$ for the background; residual gauge freedom can then be used to remove a second component, such that the potential has only transverse components $A_\LCperp \equiv A_\LCperp(x^\LCp)  \not= 0$. The nonzero part of the field strength is $F_{+\LCperp} = \partial_\LCp A_\LCperp$. This choice of potential has the advantage of making the physical (kinematic) particle momentum manifest in scattering calculations~\cite{Dinu:2012tj}.  The fields here may take any shape we choose. To emphasise that we treat the background exactly, i.e.~without recourse to perturbation in the field strength, we absorb the coupling into the field, writing $a_\mu := eA_\mu$.
%
\subsection{Helicity flip}
%
Rather than calculate the polarisation tensor we calculate directly the observables of interest, namely the scattering amplitudes for photons of momentum $l_\mu\to l'_\mu$ and helicity state $\epsilon_\mu \to \epsilon'_\mu$. (These could also be obtained by first calculating the polarisation tensor and then contracting with appropriate asymptotic states, see e.g.~\cite{Baier:1975ff,Dinu:2013gaa,Meuren:2014uia} and references therein.) However, on-shell scattering of single photons is automatically forward in a plane wave. This is a consequence of the integrability of the equations of motion;  classically, the transverse canonical momenta and the physical longitudinal momentum are conserved in the Lorentz force equation, and quantum mechanically the fermion propagator depends nontrivially only on $x^\LCp$, which leads to overall conservation of three momentum components in scattering amplitudes. 
While these conservation laws force an on-shell photon to scatter forward, internal degrees of freedom can still change. In particular, the photon helicity can flip, which is the microscopic description of vacuum birefringence~\cite{Dinu:2013gaa}.  We therefore consider the total probability of helicity-flip $\mathbb{P}_\text{flip}$, which is (with appropriately normalised wave packets taken into account, see below), just the probability for forward-scattering plus helicity-flip. 

In order to guide us we briefly mention two properties of the flip probability before beginning the calculation. First, an $S$-matrix calculation, or a calculation in lightfront perturbation theory, expresses the flip probability as a double integral~\cite{Becker-Mitter,Dinu:2013gaa,Meuren:2014uia} over two lightfront times $\phi$ and $\theta$ which are related to the spacetime vertex positions $\phi\pm \theta/2$ in the relevant Feynman diagram, see Fig.~\ref{FIG:LOOP}. Second, the integrand in the probability depends on the `effective mass'
\be\label{massa1}
	M^2 = m^2 - \langle a^2 \rangle + \langle a \rangle^2 \;,
\ee
defined by the moving average~\cite{Kibble:1975vz}
\be\label{av1}
	\langle f \rangle = \frac{1}{\theta} \int\limits^{\phi + \theta/2}_{\phi - \theta/2}\!\ud\varphi f(\varphi) \;.
\ee
The effective mass appears classically after averaging a particle's kinetic momentum $\pi_\mu$ over propagation time, $\langle \pi_\mu\rangle\langle \pi^\mu\rangle = M^2$. It appears in scattering probabilities after integrating over final states~\cite{Dinu:2013hsd}, and in scattering amplitudes after performing loop integrals~\cite{Kibble:1975vz,Dinu:2013gaa}; see e.g.~\cite{Harvey:2012ie} for a discussion of the phenomenology of the effective mass. We now proceed with our worldline calculation.
\section{Scalar QED}\label{SECT:SQED}
%
In the worldline formalism the helicity-flip amplitude in scalar QED (``sQED'') is~\cite{Schubert:2000yt}
\be\label{scalarT}
	\mathbb{T} = (ie)^2\int\limits_0^\infty\!\frac{\ud T}{T} \oint\!\mathcal{D}x^\mu e^{i S } \int\limits_0^1\!\ud \sigma'\ e^{il'.x(\sigma')} \epsilon'.\dot{x}(\sigma') \int\limits_0^1\!\ud \sigma\ e^{-il.x(\sigma)} \epsilon.\dot{x}(\sigma) \;,
\ee
where the worldline action $S$ describes a relativistic particle coupled to the background field $a_\mu$,
\be
	S = -m^2\frac{T}{2} - \int\limits_0^1\!\ud\tau\, \frac{\dot x^2}{2T} + a(x).\dot{x} \;,
\ee
and a dot is a derivative with respect to $\tau$, which parameterises the worldline. As usual (see e.g. \cite{Schubert:2000yt}) we exponentiate the vertex operators in (\ref{scalarT}) and write their contribution as a source term in the action, remembering to keep, at the end, only terms which are linear in all polarisation vectors:
\be
	\mathbb{T} = (ie)^2\int\limits_0^\infty\!\frac{\ud T}{T} \int\limits_0^1\!\ud \sigma'\ud \sigma\oint\!\mathcal{D}x^\mu \exp\bigg[i S -i \int\limits_0^1\!\ud\tau\ J_\mu x^\mu(\tau)\bigg] \;,
\ee
where the source is
\be\label{source}
	J_\mu = l_\mu \delta(\tau-\sigma) - l'_\mu \delta(\tau-\sigma') + \epsilon_\mu\dot{\delta}(\tau-\sigma) - \epsilon'_\mu\dot{\delta}(\tau-\sigma') \;.
\ee
We divide each coordinate $x$ into a centre of mass $x_c$ and an oscillating, non-constant piece $y$, so $x^\mu = x_c^\mu + y^\mu$, where (see the appendix)
\be
	\int\limits_0^1\!\ud \tau\, y^\mu(\tau) = 0 \;.
\ee
The $x^\LCperp_c$ and $x^\LCm_c$ integrals can be performed immediately to yield the momentum conservation laws described above:
\be\label{efter1}
	\mathbb{T} = \frac{(ie)^2}{2}(2\pi)^3\delta^3_{\LCperp,\LCm}({l}' - { l})\int\!\ud x^\LCp_c \int\limits_0^\infty \!\frac{\ud T}{T} \int\limits_0^1\!\ud \sigma'\ud \sigma\,  T^{2}\oint\!\mathcal{D} y^\mu
	 \exp\bigg[i (l'_\LCp - l_\LCp)x^\LCp_c + i S -i \int\limits_0^1\!\ud\tau\ J_\mu y^\mu(\tau)\bigg]
	 \;, 
\ee
where the leading $1/2$ comes from the Jacobian for the change to lightfront variables in $\ud^4x_c$.  We will perform the integrals above from right to left. Since our photon is on-shell, momentum conservation sets $l'_\LCp = l_\LCp$, i.e.~scattering is forward, and the first term in the exponent of (\ref{efter1}) drops out. 

We have that $l.\epsilon' = l.\epsilon=0$. We can simplify matters by observing that we can always make a gauge transformation such that $\epsilon^\LCp=\epsilon^{\prime\LCp} = 0$. It is not necessary to make this transformation, but it reduces the number of terms we have to consider\footnote{This is easily shown provided $l_\LCm\not=0$. If $l_\LCm=0$ then $l_\LCperp=0$ and hence $\epsilon_\LCm=0$ automatically in order to fulfil $\epsilon.l=0$; however in this case the probe photon travels parallel to the laser and there is no helicity flip, so we can ignore this case.}.
\begin{figure}[t]
\includegraphics[width=\textwidth]{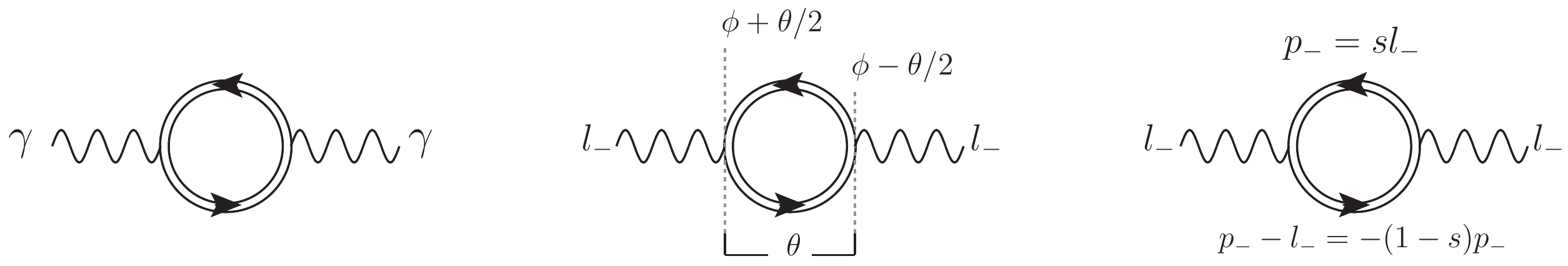}
\caption{\label{FIG:LOOP} \textit{Left:} The Furry-Feynman diagram for photon-photon scattering in a background field, to one loop. The double line represents fermions dressed (to all orders) by the background. \textit{Middle:} The two vertices have lightfront-time coordinates $x^\LCp= \phi \pm \theta/2$. \textit{Right:} In a plane wave the `minus' momentum component is conserved, as in vacuum. The momentum flowing around the loop is parameterised in terms of the momentum fraction $s := n.p/n.l = p_\LCm / l_\LCm$.}
\end{figure}
The next step is to perform the $y^\mu$ integrals. To do so we shift $y^\LCp$ and $y^\LCperp$ by a solution to the equations of motion. The classical $y^\LCp$ is found by solving
\be
	\frac{1}{T} {\ddot y}_\text{cl}^\LCp(\tau) = J^\LCp(\tau) \implies  \frac{1}{T} y_\text{cl}^\LCp(\tau) = \int\limits_{0}^1\ud \tau'\ G(\tau,\tau')J^\LCp(\tau') \;,
\ee
in which $G$ is the inverse of $\ud^2/\ud\tau^2$ on the space of periodic functions with zero average~\cite{Strassler:1992zr}, hence
\be\label{G-DEF}
	\frac{\ud^2}{\ud\tau^2} G(\tau,\tau') = \delta(\tau-\tau') -1 \;.
\ee
(For the case of constant background fields the action remains quadratic in the coordinates, and it is common to then perform calculations using the background-modified worldline propagator. Here though it suffices to use the free $G$, in both scalar and spinor QED. The reason why will become apparent below.) We will see that the two terms in (\ref{G-DEF}) correspond precisely to the two field-dependent terms in the effective mass~(\ref{massa1}). Having determined the classical $y^\LCp$ we can find the classical $y^\LCperp$:
\be
	\frac{1}{T} {\ddot y}_\text{cl}^\LCperp(\tau) = J^\LCperp(\tau) - \dot{a}^\LCperp(x^\LCp_c + y_\text{cl}^\LCp + y^\LCp) \implies  \frac{1}{T} y_\text{cl}^\LCperp(\tau) = \int\limits_{0}^1\ud \tau'\ G(\tau,\tau'){\tilde J}^\LCperp(\tau') \;,
\ee
where (recalling that only the perp components of $a$ are nonzero)
\be
	\tilde{J}_\mu = J_\mu(\tau) - \dot{a}_\mu(x^\LCp + y^\LCp_\text{cl}(\tau) + y^\LCp(\tau)) \;.
\ee
Note that we have done nothing with $y^\LCm$. Shifting $y$ by the classical solution, the amplitude $\mathbb{T}$ becomes
\be\label{efter2}
	\mathbb{T}=\frac{(ie)^2}{2}(2\pi)^3\delta^3_{\LCperp,\LCm}(l' - l) \int\!\ud x^\LCp_c\int\limits_0^\infty\!\frac{\ud T}{T} \int\limits_0^1\!\ud \sigma'\ud \sigma
	T^2\oint\!\mathcal{D} y^\mu  \exp\bigg[-i\frac{m^2T}{2} -i\frac{T}{2} \int\limits_0^1 \tilde{J}_\mu G \tilde{J}^\mu
	-i\int\limits_0^1 \frac{\dot{y}^2}{2T} + y^\LCp J_\LCp\bigg]
	\;, 
\ee
which still looks formidable due to the appearance of $y^\LCp$, which is to be integrated over, \textit{inside} the background field and therefore inside $J^\mu$ and ${\tilde J}^\mu$. However, we observe that the $y^\LCm$ integral can be evaluated exactly, as in the free theory, to give a delta functional which kills the oscillatory $y^\LCp$ in the rest of the integral. (A similar trick is applied to pair production in~\cite{Ilderton:2014mla}, see also \cite{Halpern:1976gd,Halpern:1977he}.) Hence all the $y$--integrals may be evaluated precisely, just as in the free theory, even though they are not Gaussian. The $x_c$ and $y$ integrals contribute together the free-theory factor $(2\pi T)^{-2}$, see~\eqref{pathnorm}. What remains is
%
%
\be\label{efter3}
	\mathbb{T} = -\frac{e^2}{2}(2\pi)^3\delta^3_{\LCperp,\LCm}(l' - l)\int\!\ud x^\LCp_c\int\limits_0^\infty\!\frac{\ud T}{T} (2\pi T)^{-2}\int\limits_0^1\!\ud \sigma'\ud \sigma
	 \exp\bigg[-i\frac{ m^2T}{2}  -i\frac{T}{2}\int\limits_0^1 {\tilde J}_\mu G \tilde{J}^\mu\bigg]
	 \;, 
\ee
where now $\tilde{J}_\mu = J_\mu(\tau) - \dot{a}_\mu(x^\LCp + y^\LCp_\text{cl}(\tau))$. The expression (\ref{efter3}) is almost identical to that which would be obtained in the free theory, except that the source here contains the gauge field. In fact the exponent in (\ref{efter3}) is nothing but $S[x_\text{cl}]$, that is the action evaluated on the classical trajectory obeying
\be
	\frac{1}{T}\ddot{x}_\text{cl}^\mu = J^\mu + e {F^\mu}_\nu\dot{x}_\text{cl}^\nu \;.
\ee
 In other words the semiclassical approximation is the {\it exact result} here. This is a typical property of Gaussian integrals, i.e.~of free theories or theories with {\it constant} background fields~\cite{Affleck:1981bma,Gordon:2014aba}. Here our fields are inhomogeneous but the semiclassical result is still exact; we comment further on this in Sect.~\ref{SECT:CONCS}. That the semiclassical result is exact is also a well known property of the Volkov wavefunctions which solve the Dirac and Klein-Gordon equations in a plane wave background.

We now evaluate the exponent in (\ref{efter3}) and expand the polarisation terms back out. As we are looking at a scattering amplitude there are only a few terms, and one quickly arrives at
\be\label{PRE-Z}
	\mathbb{T} = \frac{(ie)^2}{2(2\pi)^2}(2\pi)^3\delta^3({\sf l}' - {\sf l}) \int\!\ud x^\LCp_c\int\limits_0^\infty\!\frac{\ud T}{T} \mathcal{Z} \;,
\ee
in which we have defined
\be\label{Z-DEF}
	\mathcal{Z} := \int\limits_0^1\!\ud \sigma'\ud \sigma\ 
	\epsilon'.\big(\langle a\rangle - a_{\sigma'}\big)\epsilon.\big(\langle a\rangle - a_\sigma\big)
	 \exp\bigg[-i\frac{T}{2}\bigg(m^2-\langle\!\langle a^2\rangle\!\rangle+\langle\!\langle a\rangle\!\rangle^2\bigg)\bigg]\;,
\ee
with a worldline average $\langle\!\langle \cdots \rangle\!\rangle$ which arises from contractions $a\ddot{G}a$; 
\be
	\langle\!\langle a \rangle\!\rangle = \int\limits_0^1\!\ud\tau\, a\big(x^\LCp + y^\LCp_\text{cl}(\tau) \big).
\ee
The structure in the exponent of $\mathcal{Z}$ is similar to that in the effective mass (\ref{massa1}), but the average is taken over the worldline rather than a spacetime variable as in (\ref{av1}). `Projecting' the average into target space is in fact the key to performing the $\sigma$, $\sigma'$--integrals. One can show using the explicit form of $y^\LCp_\text{cl}$ that, writing $s:= |\sigma - \sigma'|$,
\be\label{aav}
	\langle\!\langle a�\rangle\!\rangle = \frac{1}{T l^\LCp s (1-s)} \int\limits^{Tl^\LCp s (1-s)/2}_{-{Tl^\LCp s (1-s)/2}}\!\ud y\, a(x^\LCp_c + y) \;.
\ee
Now change variable in (\ref{PRE-Z}) from $T$ to 
\be\label{BoV}
	\theta := T l^\LCp s (1-s) \;,
\ee
and change notation to $\phi := n.x_c = x^\LCp_c$. The worldline average then becomes exactly equal to (\ref{av1}), $\langle\!\langle \cdots \rangle\!\rangle = \langle \cdots \rangle$, and hence the exponent in $\mathcal Z$ recovers the effective mass (\ref{massa1}). In the contraction $a\ddot{G}a$ the two terms in (\ref{G-DEF}) generate the averages over $a^2$ and over $a_\mu$ squared, respectively. Without the ``background charge density'' smeared over the loop, represented by the `1' in (\ref{G-DEF})~\cite{Strassler:1992zr}, the effective mass would not be correct.  The averages over lightfront time appear in the Feynman diagram approach after performing the transverse loop momentum integrals; these integrals do not appear explicitly in the worldline approach.

The change of variable (\ref{BoV}) maps the intrinsic length of the worldline path, $T$, onto  $\theta$, the spacetime separation between Feynman diagram vertices in the $n.x$~direction. (This could have been guessed at the beginning: both $\theta$ and $T$ are integrals over the positive real line.) When rewritten in terms of the new variables the terms outside the exponent  in (\ref{Z-DEF}) become
\be
	a(x_c^\LCp + y^\LCp_\text{cl}(\sigma')) = a ( \phi + \theta/2) \;, \qquad 	a(x_c^\LCp + y^\LCp_\text{cl}(\sigma)) = a ( \phi - \theta/2) \;,
\ee
independent of $\sigma$, $\sigma'$, the integrals over which can now be performed. Because the integrand depends on $\{\sigma,\sigma'\}$ only in the combination $s(1-s)$, one can show directly that there remains no dependence on~$\sigma'$ after evaluation of the $\sigma$ integral.  This is just as in the theory without background~\cite{Strassler:1992zr}. Hence we drop the $\sigma'$ integral (which contributes a factor of unity) and integrate over $\sigma\to s$. Writing $x \equiv  M^2/n.l$ this integral is, with $K_j$ the modified Bessel functions,
\be\label{I-1}
	\int\limits_0^1\!\ud s\, \exp\bigg(-i \frac{x}{2s(1-s)}\bigg) =  i x e^{-i x } \big(K_1(ix) - K_0(ix) \big)=: \mathcal{I}_1(x) \;.
\ee
This integral over the relative vertex position $\sigma-\sigma'$ appears in the Feynman diagram and lightfront approaches as an integral over the {\it lightfront momentum fraction} $s = n.p/n.l = p_\LCm / l_\LCm$, where $p$ is the momentum of the fermion in the loop, see Fig.~\ref{FIG:LOOP}, right hand panel. In those approaches the integral limits arise from momentum conservation, but here they were in place from the beginning. The appearance of the ``$s(1-s)$'' factors is also typical of lightfront wavefunctions~\cite{Lepage,Brodsky:1997de,Heinzl:2000ht}.

Collecting factors, we find the near-final result
\be
	\mathbb{T} = -(2\pi)^3 \delta_{\LCperp,\LCm}(l'-l) \frac{\alpha}{2\pi}  \int\!\ud \phi \int\limits_0^\infty\!\frac{\ud \theta}{\theta}   \mathcal{I}_1\big(\tfrac{\theta M^2}{n.l}\big) \epsilon'.\big(a(\phi +\theta/2)-\langle a \rangle\big)\epsilon.\big(a(\phi -\theta/2)- \langle a \rangle\big) \;.
\ee
To compactify notation we define
\be
	{\sf A}:= \epsilon.\langle a \rangle \;, \quad \bar{\sf A}:= \epsilon'.\langle a \rangle \;, \quad  {\sf A}_\theta := \partial_\theta {\sf A}\,;  \quad \quad  {\sf A}_\phi := \partial_\phi {\sf A}\;, \quad\text{and so on.}
\ee
With this and the results
\be
	\pm \epsilon.\big(a(\phi \pm \theta/2)- \langle a \rangle \big) = \theta \big(\tfrac{1}{2}{\sf A}_\phi \pm {\sf A}_\theta \big) \;,
\ee
we can write our amplitude in the form
\be\label{reducerad}
	\mathbb{T} = -(2\pi)^3 \delta_{\LCperp,\LCm}(l-l')\, n.l \, \mathcal{M}_\text{scalar}.
\ee
To obtain the probability we take the modulus squared of (\ref{reducerad}) and integrate over final states. This integral converts one factor of $(2\pi)^3\delta^3$ into a $1/n.l$, while the second delta function is regularised by including a photon wavepacket from the beginning, and gives a second factor of $1/n.l$, see~\cite{Ilderton:2012qe}. Hence we obtain $\mathbb{P}_\text{flip} = |\mathcal{M}_\text{flip}|^2$ or, explicitly,
\be\label{P-SKAL�R}
\begin{split}
	\mathbb{P}_\text{flip,scalar} = \bigg| \frac{\alpha}{2\pi}\frac{1}{n.l}\int\limits_{-\infty}^\infty\!\ud \phi \!\int\limits_0^\infty\!\ud\theta\theta\   \mathcal{I}_1\big(\tfrac{\theta M^2}{n.l}\big)&(\bar{\sf A}_\theta+\tfrac{1}{2}\bar{\sf A}_\phi)({\sf A}_\theta-\tfrac{1}{2}{\sf A}_\phi) \bigg|^2 \;,
\end{split}
\ee
which is rather compact, and of the same form as the spinor QED result found in~\cite{Dinu:2013gaa} using lightfront methods. The two remaining integrals over the lightfront times $\phi$ and $\theta$ can be performed analytically only for special cases (but then one must give up performing the $s$-integral exactly~\cite{RitusReview}), and in general must be tackled numerically. For examples and details of the numerical method of integration see~\cite{Dinu:2013gaa}.
\subsection{Analyticity}
In evaluating the contraction $\tilde{J}^\mu G \tilde{J}_\mu$ in (\ref{efter3}) to obtain (\ref{Z-DEF}) there arises a term
	 \be
	 	\int\limits_0^1\!\ud \tau\; \dot{y}^\LCp_\text{cl}\, a(x^\LCp_c + y^\LCp_\text{cl}) \;.
	 \ee
 This integral is exact and, by the boundary conditions, vanishes, provided that the potential $a$ is an integrable function without singularities. This is of course natural, and we remark only that similar terms appear in worldline instanton calculations of pair production, see e.g.~\cite{Ilderton:2015qda} and the discussion of the argument principle in~\cite{Ilderton:2015lsa}.

\subsection{Zero modes}
As our results parallel those found using lightfront quantisation we should comment on the appearance of lightfront zero modes in this formalism. Zero modes are states for which the momentum component $p_\LCm=0$, and there is a vast literature on the role they play and the problems they can cause through the appearance of $1/p_\LCm$ factors, see~\cite{Brodsky:1997de,Heinzl:2000ht} and references therein. For this discussion it is enough to check that the zero modes do not introduce divergences into our calculations~\cite{Dinu:2013gaa,Meuren:2014uia}.

The change of variable (\ref{BoV}) may look suspect when $s\in\{0,1\}$, that is when $\sigma=\sigma'$, $\{\sigma=1, \sigma'=0\}$, or $\{\sigma'=1,\sigma=0\}$. For  these values the vertex operators lie at the same position on the worldline. Now, we have already identified that $s$ corresponds to the momentum fraction $p_\LCm/l_\LCm$ flowing around the loop. From the right-hand diagram in Fig.~\ref{FIG:LOOP} this means that $s=0,1$ corresponds to a zero mode flowing through the upper or lower portion of the loop, respectively. We observe that $y^\LCp_\text{cl} \to 0$ at these points, which implies that the background field vanishes from $\tilde{J}$ in \eqref{efter3}, and we return to the free theory, where the helicity-flip amplitude must vanish. This suggests that we should not expect problems from such points. Indeed it is seen directly that there are no singularities in (\ref{Z-DEF}) as $s\to\{0,1\}$, so the zero modes do not cause problems. Also as expected, using that $y^\LCp_\text{cl} \to 0$ and that $\langle a_\mu \rangle \to a_\mu(x^\LCp_c) $ from (\ref{aav}), the vertex operator insertions vanish in (\ref{Z-DEF}) so that there is no contribution to the flip probability from the zero modes. 
%
%
\section{Spinor QED}\label{SECT:QED}
%
We now turn to the helicity-flip amplitude in spinor QED. Spin contributions are included by inserting a ``spin factor'' into the scalar worldline integral, of form \cite{Feynman:1951gn,Schubert:2001he,CS}
	\be\label{Spin}
\text{Spin}=\frac{1}{4}\text{tr }P\exp\bigg[-\frac{iT}{4}\int\limits_0^1\ud\tau\sigma^{\mu\nu}\mathcal{F}_{\mu\nu} \bigg]\;,
\ee
where the fieldstrength $\mathcal{F}_{\mu\nu}$ contains both the background and scattered photons (definition below), $\sigma^{\mu\nu}=\frac{i}{2}[\gamma^\mu,\gamma^\nu]$, the trace is over the Dirac matrices, and $P$ stands for path-ordering. (We define $\text{Spin}$ with an extra factor of $1/4$ compared to~\cite{CS}, and we work in Minkowski rather than Euclidean space.) It is sometimes more convenient to rewrite $\text{Spin}$ as a path integral over Grassmann variables on the worldline; doing so makes explicit a worldline supersymmetry~\cite{Brink:1976sz}, the use of which in worldline calculations is reviewed in~\cite{Schubert:2001he}.  Another method for evaluating the Grassmann integral, for arbitrary background fields, is presented in~\cite{Gitman:1996wm}. There the Grassmann fields are integrated out without approximation to obtain a worldline path integral for spinor QED which depends only on the worldline coordinates, as for sQED. Here we will first consider the form~\eqref{Spin}, since only a few terms in the expansion of the exponent are nonzero, but we will also use the Grassmann approach, below. (The form (\ref{Spin}) was also used in~\cite{Dunne:2005sx} to obtain the spinor contribution to pair production.) 

\subsection{Feynman's Dirac-trace approach}
The helicity flip amplitude in QED is
\be\label{amp1}
	\mathbb{T}_\text{flip} =2e^2\int\limits_0^\infty\frac{\ud T}{T}\oint\mathcal{D}x\exp\bigg[-im^2\frac{T}{2}-i\int\limits_0^1\!\ud\tau\ \frac{\dot{x}^2}{2T} + \mathcal{A}.\dot{x}\bigg]\, \text{Spin} \;,
\ee
where $\mathcal{A}_\mu=a_\mu+\epsilon_\mu e^{-il.x}+\epsilon'_\mu e^{il'.x}$, and the fieldstrength $\mathcal{F}$ in the spin factor is $\mathcal{F}_{\mu\nu}=\partial_\mu \mathcal{A}_\nu-\partial_\nu \mathcal{A}_\mu$. As before we retain only those terms which are linear in both $\epsilon$ and $\epsilon'$.  Note that (\ref{amp1}) is obtained from the sQED expression~(\ref{scalarT}) by inserting the spin-factor (\ref{Spin}) and multiplying by two.

We begin by expanding out the exponential in the spin factor, using
\be
-\frac{i}{2}\sigma^{\mu\nu}\mathcal{F}_{\mu\nu}=\slashed{n}\slashed{a}'-i\slashed{l}\slashed{\epsilon}e^{-il.x}+i\slashed{l}'\slashed{\epsilon}'e^{il'.x} \;.
\ee
Only a few of the terms are nonzero because of our chosen background and process. An analogous truncation of terms occurs when one solves the Lorentz force equation in a plane wave: a path-ordered exponential truncates at second order because the field strength $F_{\mu\nu}$ is nilpotent of order three.

When selecting the terms that are linear in $\epsilon$ and $\epsilon'$ it is convenient to separate the total amplitude into three parts, writing $\mathbb{T}_\text{flip}=\mathbb{T}_0+\mathbb{T}_1+\mathbb{T}_2$. The term $\mathbb{T}_0$ is identical to the amplitude in sQED up to the factor of $2$, and receives both $\epsilon$ and $\epsilon'$ from the $\mathcal{A}.\dot{x}$ term in \eqref{amp1}. The term $\mathbb{T}_1$ receives one of the polarisation vectors from $\mathcal{A}.\dot{x}$ and the other from $\text{Spin}$. The term $\mathbb{T}_2$ receives both $\epsilon$ and $\epsilon'$ from $\text{Spin}$. Writing arguments as subscripts to compactify notation, the spinor contribution to $\mathbb{T}_1$ is obtained from
\be\label{trace1}
	\frac{1}{4}\text{tr}\int\limits_0^1\!\ud\tau_2\ud\tau_1\; \theta(\tau_2-\tau_1)\frac{T^2}{4}(\slashed{n}\slashed{a}'-i\slashed{l}\slashed{\epsilon}e^{-il.x}+i\slashed{l}'\slashed{\epsilon}'e^{il'.x})_{\tau_2}(\slashed{n}\slashed{a}'-i\slashed{l}\slashed{\epsilon}e^{-il.x}+i\slashed{l}'\slashed{\epsilon}'e^{il'.x})_{\tau_1} \;,
\ee
and the spinor contribution to $\mathbb{T}_2$ is obtained from
\be
	\frac{1}{4}\text{tr} \int\limits_0^1\!\ud\tau_{4321}\; \theta_{4321}\frac{T^4}{16}(\slashed{n}\slashed{a}'-i\slashed{l}\slashed{\epsilon}e^{-il.x}+i\slashed{l}'\slashed{\epsilon}'e^{il'.x})_{\tau_4}(...)_{\tau_3}(...)_{\tau_2}(...)_{\tau_1} \;,
\ee
where $\theta_{4321}=\theta(\tau_4-\tau_3)\theta(\tau_3-\tau_2)\theta(\tau_2-\tau_1)$. After performing the traces the three `partial amplitudes' $\mathbb{T}_i$ can be written in terms of the same current $J$ as in (\ref{source}) as 
\be\label{Gamma-Ii}
	\mathbb{T}_i=2e^2\int\limits_0^\infty\frac{\ud T}{T}e^{-iTm^2/2}\oint\mathcal{D}x \; I_i\exp\bigg[-i\int\limits_0^1\frac{\dot{x}^2}{2T}+a.\dot{x}+J.x \bigg]\;,
\ee
where the three integrands $I_i$ are, writing ``$\text{lin}_\epsilon$'' for the instruction to select the term linear in $\epsilon$,
\bea
	I_0 &=& -\int\limits_0^1\ud\sigma\ud\sigma' \; \text{lin}_\epsilon\text{lin}_{\epsilon'} \;, \\
	I_1 &=& \frac{iT^2}{4}\int\limits_0^1\!\ud\sigma\ud\sigma' \;\Big(n.l \int\limits_0^1\!\ud\tau \; \epsilon.a' \; \text{lin}_{\epsilon'}+n.l'\int\limits_0^1\!\ud\tau\; \epsilon'.a' \; \text{lin}_\epsilon\Big)  \;, \\
\label{I2}
	I_2 &=& n.l n.l'\frac{T^4}{8}\int\limits_0^1\ud\tau_{4321}\Big|_{\tau_3=\sigma,\tau_1=\sigma'}(\theta_{4321}+\theta_{2143}+\theta_{3214}+\theta_{1432})(\epsilon.\underset{4}{a'}\epsilon'.\underset{2}{a'}+\epsilon.\underset{2}{a'}\epsilon'.\underset{4}{a'}) \;.
\eea
All explicit reference to spinors has now been eliminated, and we are left with a worldline integral over the coordinates, as in~\cite{Gitman:1996wm}.  The integral can be performed with the same method as for the scalar case, above.  $\mathbb{T}_0$ is the same as in sQED up to a factor of $2$:
\be\label{Gamma0}
	\mathbb{T}_0=-(2\pi)^3\delta_{\LCm,\LCperp}(l'-l)\frac{\alpha}{\pi}
	\int\!\ud \phi \!
	\int\limits_0^\infty\ud\theta \; \theta\!
	\int\limits_0^1\!\ud s\;
	\big( \bar{\mathsf{A}}_\theta+\tfrac{1}{2} \bar{\mathsf{A}}_\phi \big)
	\big(\mathsf{A}_\theta-\tfrac{1}{2}\mathsf{A}_\phi \big)e^{-\frac{i\theta M^2}{2n.ps(1-s)}} \;.
\ee
For $\mathbb{T}_1$ we use
\be
	\int\limits_0^1\!\ud\tau\;  f'(x^\LCp+y_\text{cl}^\LCp(\tau;\sigma,\sigma')) = \frac{1}{\theta}\int\limits_{\phi-\theta/2}^{\phi+\theta/2}\ud\varphi \; f'(\varphi) = \langle f\rangle_\phi \;,
\ee
and that the integrand again depends only on the combination $s(1-s)$ with $s=|\sigma-\sigma'|$. This gives
%
%
\be\label{Gamma1}
	\mathbb{T}_1 = -(2\pi)^3\delta_{\LCm,\LCperp}(l'-l)\frac{\alpha}{4\pi}
	\int\!\ud\phi \!
	\int\limits_0^\infty\!\ud\theta \; \theta\!
	\int\limits_0^1\!\frac{\ud s}{s(1-s)} \;
	\Big(-\bar{\mathsf{A}}_{[\phi} {\mathsf{A}}_{\theta]} + \bar{\mathsf{A}}_{\phi} {\mathsf{A}}_{\phi}\Big)e^{-\frac{i\theta M^2}{2n.ps(1-s)}} \;.
\ee
The term $\mathbb{T}_2$ is more complicated. By using the symmetry $\tau_4\leftrightarrow\tau_2$ and $\tau_3\leftrightarrow\tau_1$ ($\sigma\leftrightarrow\sigma'$) we can write
\be
	I_2 = n.l^2 \frac{T^4}{4}
	\int\limits_0^1\!\ud\tau_{4321}\;
	(\theta_{4321}+\theta_{3214}) \big(\epsilon.\underset{4}{a'}\epsilon'.\underset{2}{a'}+(\epsilon\leftrightarrow\epsilon')\big) \;.
\ee
Next we use the explicit solution $y_\text{cl}^\LCp(\tau;\sigma,\sigma')$ (recall $\tau_3=\sigma$ and $\tau_1=\sigma'$) to perform the integrals over $\tau_2$ and $\tau_4$,
\be
\int\limits_{\tau_1}^{\tau_3}\ud\tau_2 \; a'(x^\LCp+y_\text{cl}^\LCp)=s\langle a\rangle_\phi \;, \qquad \int\limits_0^{\tau_1}\ud\tau_4 \; a'(x^\LCp+y_\text{cl}^\LCp)+\int\limits_{\tau_3}^1\ud\tau_4 \; a'(x^\LCp+y_\text{cl}^\LCp)=(1-s)\langle a\rangle_\phi \;.
\ee
The integrand for the remaining integrals over $\sigma$ and $\sigma'$ again only depends on $s(1-s)$, except for a factor of $\theta_{31}$ which simply gives a factor of $1/2$. The final result for $\mathbb{T}_2$ is
\be\label{Gamma2}
	\mathbb{T}_2 = (2\pi)^3\delta_{\LCm,\LCperp}(l'-l)\frac{\alpha}{4\pi}
	\int\!\ud\phi\!
	\int\limits_0^\infty\!\ud\theta \; \theta\!
	\int\limits_0^1\! \frac{\ud s}{s(1-s)}\; \bar{\mathsf{A}}_{\phi} {\mathsf{A}}_{\phi} \, e^{-\frac{i\theta M^2}{2n.ps(1-s)}} \;.
\ee
Note that~\eqref{Gamma2} is cancelled entirely by a similar term in~\eqref{Gamma1}. The integrals over $s$ give either $\mathcal{I}_1$ introduced in~(\ref{I-1}) or
\be\label{I-2}
	\mathcal{I}_2(x) :=  \int\limits_0^1\!\frac{\ud s}{2 s(1-s)}\, \exp\bigg(-i \frac{x}{2s(1-s)}\bigg) =  e^{-i x }K_0(ix) \;.
\ee
Adding the three contributions \eqref{Gamma0}, \eqref{Gamma1} and \eqref{Gamma2} together we find
%
%
%
\be\label{spinor-final}\begin{split}
	\mathbb{T}_\text{flip} &=
	-(2\pi)^3\delta_{\LCm,\LCperp}(l'-l)
	\frac{\alpha}{\pi}
	\int\!\ud\phi \int\limits_0^\infty\!\ud\theta \; \theta\bigg[
	(\bar{\sf A}_\theta+\tfrac{1}{2}\bar{\sf A}_\phi)({\sf A}_\theta-\tfrac{1}{2}{\sf A}_\phi) 
\mathcal{I}_1 - \frac{1}{2}\bar{\mathsf{A}}_{[\phi} \mathsf{A}_{\theta]} \mathcal{I}_2 \bigg] \\
& = -(2\pi)^3 \delta_{\LCperp,\LCm}(l-l')\, n.l \, \mathcal{M}_\text{spinor} \;,
\end{split}\ee
in exact agreement with the results in~\cite{Dinu:2013gaa} found using lightfront quantisation and standard Feynman diagram methods. 
Squaring the amplitude yields, as above, $\mathbb{P} = |\mathcal{M}|^2$, with
\be\label{P-SPINOR}
\begin{split}
	\mathbb{P}_\text{flip,spinor} = \bigg| \frac{\alpha}{\pi}\frac{1}{n.l}\int\limits_{-\infty}^\infty\!\ud \phi \!\int\limits_0^\infty\!\ud\theta\theta\  \bigg( \mathcal{I}_1\big(\tfrac{\theta M^2}{n.l}\big)&(\bar{\sf A}_\theta+\tfrac{1}{2}\bar{\sf A}_\phi)({\sf A}_\theta-\tfrac{1}{2}{\sf A}_\phi) 
 - \frac{1}{2}\mathcal{I}_2 \big(\tfrac{\theta M^2}{n.l}\big) \bar{\sf A}_{[\phi} {\sf A}_{\theta]})\bigg)
 \bigg|^2 \;.
\end{split}
\ee
\subsection{Grassmann aproach}
The spin factor \eqref{Spin} can be expressed as a path integral over anticommuting Grassmann variables $\psi^\mu$~\cite{CS} (replace $T$ in that paper by $iT/2$ to go to our conventions),
\be
	\text{Spin}=\frac{1}{4}\int\mathcal{D}\psi\exp\bigg[-\frac{1}{2}\int\limits_0^1 \psi_\mu\dot{\psi}^\mu+T\psi^\mu \mathcal{F}_{\mu\nu}\psi^\nu \bigg]\;,
\ee
and the integral is calculated with anti-periodic boundary conditions $\psi(1)=-\psi(0)$. We begin by expanding the exponential in the polarisation vectors. This gives three terms, as above. The next step is to remove the gauge field from the action via the change of variable
\be\label{BYTA}
	\psi^\mu(\tau) \to \psi^\mu(\tau) - T n^\mu\int\limits_0^1\!\ud\tau' \; G_F(\tau-\tau') a'.\psi \;, \qquad G_F(\tau-\tau')=\frac{1}{2}\text{sign}(\tau-\tau')\;.
\ee
Note that only $\psi^\LCm$ is changed, and that only $\psi^\LCperp$ appears under the integral in (\ref{BYTA}). After this change of variable the exponent appearing in our amplitude becomes that of the free theory:
%
%
\be\label{curly}\begin{split}
\text{Spin}&=\frac{1}{4} \int\!\mathcal{D}\psi\;\bigg\{
1 - iT \mathcal{J}-\frac{1}{2}T^2\mathcal{J}^2\bigg\}
	\exp \bigg[-\frac{1}{2}\int\limits_0^1\! \psi_\mu\dot{\psi}^\mu \bigg]  \;,\\
	\mathcal{J} &:=\int\limits_0^1\!\ud\tau\ud\tau' 
	\big(\psi^\mu(\tau)-Tn^\mu G_F(\tau-\tau')a'.\psi(\tau')\big)
	\left(l'_\mu\epsilon'_\nu e^{il'.x(\tau)} -l_\mu\epsilon_\nu e^{-il.x(\tau)}\right)\psi^\nu(\tau) \;.
\end{split}
\ee
(The same change of variables in the effective action would remove all dependence on the field, turning it into the  effective action of the free theory; this is because there is no Schwinger pair production in a plane wave.) The Grassmann integral can now be performed using the same Wick contractions as in the free theory~\cite{CS}. This is trivial for the first term in~(\ref{curly}), and a straightforward calculation shows that the second term agrees with that obtained by performing the trace in~\eqref{trace1}. The third term in~(\ref{curly}) requires a longer calculation. At an intermediate step, one finds
\be\label{inter}
	\frac{T^4 n.l^2}{16}\int\limits_0^1\!\ud\tau_{4321} \; {e^{-il.x(\tau_3)}} {e^{il'.x(\tau_1)}} \epsilon.\underset{2}{a'}\epsilon'.\underset{4}{a'}(1-16G^F_{43}G^F_{32}G^F_{21}G^F_{14}) \;.
\ee
By writing $G^F_{ij}=\theta_{ij}-\theta_{ji}$ and using that $\theta_{ij}+\theta_{ji}=1$ 
one can show that~\eqref{inter} is equal to~\eqref{I2}, and one recovers~\eqref{Gamma-Ii}. The rest of the calculation is identical to that above.
%
%
\section{Discussion and conclusions}\label{SECT:CONCS}
%
We have derived the photon helicity-flip probability in inhomogeneous plane wave backgrounds using the  worldline formalism. The calculation was performed to one loop, but exactly in all other parameters.  The calculation in scalar QED is direct and the final result compact. Both this and the QED calculation recover the expressions previously found using lightfront methods, thus our results also serve to confirm the advantages of this approach to strong field QED~\cite{Bakker:2013cea}.

The sQED calculation does seem simpler in the worldline formalism than
in approaches based on the Volkov solution, and so offers a promising tool for the calculation of other processes, including for example multi-photon emission: by combining the methods presented here with those in~\cite{Ahmadiniaz:2015xoa,Ahmadiniaz:2015kfq} is should be possible to obtain useful expressions for the $n$-photon emission amplitudes, for arbitrarily high~$n$. It is fair to say that the calculation of spin contributions is involved. Nevertheless, we have seen that two different methods can be applied, and one or the other may offer simplifications depending on the process in question.  For the problem here, using the Dirac trace is simpler than using a Grassmann integral: there are fewer steps with the former, even though the calculations as a whole are very similar.  



We have seen that the semiclassical approximation to our worldline integrals remains exact, even though the background is inhomogeneous. This provides a simple recipe for the calculation of other processes: 1) exponentiate the vertex operators to obtain the source, 2) evaluate the classical action on the classical path, 3) re-expand the vertex operators. Finally, 4) compute as many integrals as possible. The calculation of the classical path will go through as above with the appropriate source and propagator: one first solves for $x^\LCp$, and then for $x^\LCperp$. This method should hold for all processes in plane wave backgrounds.

\acknowledgments
A.~I.~thanks N.~Ahmadiniaz, S.~Meuren and C.~Schubert for many useful discussions. A.~I.~is supported by the Olle Engkvist Foundation, grant  2014/744.

\appendix

\section{Conventions}
The reparameterisation-invariant measure over the worldline coordinates is~\cite{Polyakov:1987ez,Mansfield:1990tu}
\be
	\oint\!\mathcal{D} x = \sqrt{T}\int\!\ud x_c \oint\!\mathcal{D} y \;, 
\ee
in which $x_c$ is the centre-of-mass of the loop, $y$ is the nonconstant and oscillatory part of the closed path, and the factor of $\sqrt{T}$ arises from changing variable from the Fourier zero mode of the loop to $x_c$. In four dimensions the measure obeys
%
%
\be\label{pathnorm}
	\oint\!\mathcal{D} x \exp\bigg[-\frac{i}{2T}\int\limits_0^1\!\ud\tau\ \dot{x}^2\bigg] = (2\pi T)^{-2} \int\!\ud^4 x_c \;,
\ee
with the final integral giving the spacetime volume.
The fermionic path integral measure is, in four dimensions, normalised to~\cite{CS}
\be\label{free-Grassmann}
	\int\!\mathcal{D}\psi\exp\bigg[-\frac{1}{2}\int\limits_0^1\!\ud\tau\ \psi_\mu\dot{\psi}^\mu \bigg]=4 \;.
\ee
and is calculated with anti-periodic boundary conditions $\psi(1)=-\psi(0)$.

\end{document}